\documentclass[a4wide,11pt,english]{article}
\pdfoutput=1
\usepackage[T1]{fontenc}
\usepackage[utf8]{inputenc}
\usepackage{geometry}
\usepackage{amssymb}
\usepackage{graphicx}
\usepackage{subcaption}
\usepackage{graphics}
\usepackage{amsmath}
\numberwithin{equation}{section}

\usepackage{chngcntr}
\counterwithin{figure}{section}
\usepackage{esint}
\usepackage{cite}
\usepackage{amsmath}
\usepackage{feynmp}
\usepackage{mathrsfs}
\usepackage{braket}
\usepackage{dsfont}
\usepackage{cancel}
\usepackage[utf8]{inputenc}
\usepackage{color}

\makeatletter
\providecommand{\tabularnewline}{\\}
\makeatother
\usepackage{babel}
\usepackage{hyperref}
\def\equationautorefname~#1\null{(#1)\null}
\usepackage{feynmp}
\makeatletter
\def\endfmffile{%
  \fmfcmd{\p@rcent\space the end.^^J%
          end.^^J%
          endinput;}%
  \if@fmfio
    \immediate\closeout\@outfmf							%to get the Feynman diagrams included run in shell pdflatex - shell - escape twice!!!
  \fi
  \ifnum\pdfshellescape=\@ne
    \immediate\write18{mpost \thefmffile}%
  \fi}
\makeatother
\begin{document}

\newcommand{\lij}{$\lambda_{ij}$}
\newcommand{\lam}[1]{$D_{\lambda,#1#1}$}
\newcommand{\mchi}{$m_\chi$}
\newcommand{\mphi}{$m_\phi$}
\newcommand{\tet}[1]{$\theta_{#1}$}
\newcommand{\del}[1]{$\delta_{#1}$}

%%%%%%%%%% Title page
\begin{titlepage}
\begin{flushright}
TTP17-008 
\end{flushright}
\vskip1.2cm
\begin{center}
{\LARGE \bf \boldmath Top-Flavoured Dark Matter \vspace{1mm}\\in Dark Minimal Flavour Violation}
\vskip1.0cm
{\bf \large
Monika Blanke, Simon Kast}
\vskip0.3cm
 {Institut f\"ur Kernphysik, Karlsruhe Institute of Technology,\\
  Hermann-von-Helmholtz-Platz 1,
  D-76344 Eggenstein-Leopoldshafen, Germany}\vspace{1mm}\\
 {Institut f\"ur Theoretische Teilchenphysik,
  Karlsruhe Institute of Technology, \\
Engesserstra\ss e 7,
  D-76128 Karlsruhe, Germany}

\vskip0.51cm

%{\em Version of \today}

\vskip0.35cm

{\large\bf Abstract\\[10pt]} \parbox[t]{.9\textwidth}{
We study a simplified model of top-flavoured dark matter in the framework of Dark Minimal Flavour Violation. In this setup the coupling of the dark matter flavour triplet to right-handed up-type quarks constitutes the only new source of flavour and CP violation. The parameter space of the model is restricted by LHC searches with missing energy final states, by neutral $D$ meson mixing data, by the observed dark matter relic abundance, and by the absence of signal in direct detection experiments. We consider all of these constraints in turn, studying their implications for the allowed parameter space. Imposing the mass limits {and coupling benchmarks} from collider searches, we then conduct a combined analysis of all the other constraints, revealing {their non-trivial interplay}. %among them. 
%Only when considering them simultaneously, the full extent of the exclusion bounds is revealed. 
Especially interesting is the combination of direct detection and relic abundance constraints, having a severe impact on the structure of the dark matter coupling matrix. 
We point out that future bounds from upcoming direct detection experiments, such as XENON1T, XENONnT, LUX-ZEPLIN, and DARWIN, will exclude a large part of the parameter space and push the DM mass to higher values.

%We point out that {\bf CONTINUE WITH MAIN CONCLUSIONS}
}

\end{center}
\end{titlepage}

%\maketitle

%\begin{abstract}

%\end{abstract}

%\newpage 
\setcounter{tocdepth}{2}
\tableofcontents{}

%\newpage 

\section{Introduction}

The evidence for dark matter (DM) collected in the past decades is overwhelming. Astronomical observations of velocity curves of stars in galaxies, as well as galaxy movements in clusters and gravitational lensing effects demand the presence of hidden sources of gravitational interactions \cite{Gorenstein:2014iba,Gelmini:2015zpa,Garrett:2010hd}, with new particles being the best explanation for these observations \cite{Bertone:2004pz}. The presence of DM can not only account for astronomical observations, but also provides an elegant mechanism to explain the enhanced structure formation in the early universe \cite{White:1977jf}. Furthermore a possible interaction with SM particles (aside from gravity) serves to explain the similar orders of magnitude of baryonic ($\approx 4.9\%$) and dark matter ($\approx 26.8\%$) content in the Universe \cite{Ade:2015xua}. 

WIMP (weakly interacting massive particle) candidates for DM are particularly appealing from the theoretical point of view, as these models can provide a connection between two longstanding problems of particle physics -- the origin of electroweak symmetry breaking, and the origin of DM. Further support is provided by the observation that a weak scale particle with weak annihilation cross-section straightforwardly provides the correct relic abundance by means of a thermal freeze-out, known as the WIMP miracle. 
However in spite of their conceptual beauty, so far no experimental evidence for WIMPs has been found. The increasingly stringent constraints from direct DM detection experiments and LHC searches put simple realizations of the WIMP paradigm under severe pressure.

A possible solution to this tension is provided by the idea of flavoured DM (FDM), which proposes a flavour structure in the dark sector \cite{Kilic:2015vka,Agrawal:2011ze,Cheung:2011zza,Kile:2011mn,Batell:2011tc,Kamenik:2011nb,Kumar:2013hfa,Chang:2013oia,Kile:2013ola,Bai:2013iqa,Batell:2013zwa,Agrawal:2014ufa,Agrawal:2014una,Gomez:2014lva,DMFVPrimer,Hamze:2014wca,Lee:2014rba,Kile:2014jea,Agrawal:2015kje,Lopez-Honorez:2013wla}. Such a flavour structure of DM allows in general for a non-trivial coupling to the SM flavour triplets of quarks or leptons. Consequently, the stringent constraints from direct detection experiments and LHC searches can be partially evaded. In addition, flavour symmetries can provide a stabilization mechanism for DM \cite{Batell:2011tc,DMFVPrimer}. In general, a flavour violating coupling constitutes a new source of both flavour and CP violation and yields new contributions to precision flavour observables. In order to not spoil the good agreement of the latter with their SM predictions, in most studies Minimal Flavour Violation (MFV) \cite{Buras:2000dm,D'Ambrosio:2002ex,Buras:2003jf} was imposed. 

More recently, however, the phenomenology of FDM models beyond MFV has been studied. The authors of \cite{DMFVPrimer} proposed Dark Minimal Flavour Violation (DMFV) as a minimal non-MFV framework. In DMFV,  the DM coupling to quarks constitutes the only new source of flavour and CP violation beyond the SM. The flavour phenomenology is therefore significantly altered and the constraints from precision flavour data have to be taken into account. At the same time however the number of new parameters is limited and the stability of DM remains intact. As a concrete example the DMFV hypothesis was applied to a simplified model with fermionic DM coupling to the right-handed down-type quarks via a scalar mediator. Subsequently, in \cite{Chen:2015jkt} a simplified model of lepton-flavoured DM in the DMFV framework was considered. Note that in this case an additional symmetry is required to stabilize DM.

In this paper we use the DMFV hypothesis to construct a simplified model of top-flavoured DM.   In \autoref{modelsection} we present the model and revisit in short the concept of DMFV. In \autoref{ColliderSection} we study the impact of collider searches on our model, and choose parameter benchmarks for our subsequent studies. Constraints from flavour experiments, the observed relic abundance, and direct detection experiments are discussed in \autoref{Fsection}, \autoref{RAsection}, and \autoref{DDsection}, respectively. Then, in \autoref{combinedsection} we discuss the combined effect of all these constraints on the parameter space of the model. Finally in \autoref{summarysection} we recapitulate our findings, and contemplate the prospects of future experiments for the considered model.

\section{Top-Flavoured Dark Matter beyond MFV}
\label{modelsection}

In this section we present the model analysed in the rest of this paper. It is constructed in analogy to the model coupling to down-quark discussed in \cite{DMFVPrimer}. We will revisit the most important features of a simplified model of that kind.

\subsection{DMFV: Simplified Model and Parametrization}

%Following the principles of Dark Minimal Flavour Violation (DMFV), coupling the DM flavour triplet (transforming under a new flavour symmetry $U(3)_{\chi}$, extending the global flavour symmetry) to the right-handed SM up-quark triplet, we find the most general renormalizable Lagrangian to be
%
Following \cite{DMFVPrimer}, we study a simplified model of flavoured DM with the following Lagrangian:
\begin{eqnarray}
\mathcal{L}& = &\mathcal{L}_\text{SM}+i\bar{\chi}\hspace{-1mm}\not{\hspace{-.6mm}\partial}\chi-m_{\chi}\bar{\chi}\chi-\left(\lambda_{ij}\bar{u}_{Ri}\chi_{j}\phi+h.c.\right) \\
           &   & +\left(D_{\mu}\phi\right)^{\dagger}\left(D^{\mu}\phi\right)-m_{\phi}^2\phi^{\dagger}\phi+\lambda_{H\phi}\phi^{\dagger}\phi H^{\dagger}H+\lambda_{\phi\phi}\left(\phi^{\dagger}\phi\right)^{2}\,. \nonumber
\end{eqnarray}
Here, the field $\chi$ is a Dirac fermion which is a singlet under the SM gauge group, and it transforms as a triplet under a global $U(3)_\chi$ flavour symmetry. Its lightest flavour constitutes the observed DM. $\chi$ couples to the SM up-quarks via a scalar mediator $\phi$ carrying QCD colour and hypercharge. Employing the Dark Minimal Flavour Violation (DMFV) paradigm \cite{DMFVPrimer}, the quark-DM coupling matrix $\lambda$ is a general 3$\times$3 complex matrix, which is assumed to be the only new source of both flavour and CP violation. The couplings $\lambda_{H\phi}$ and $\lambda_{\phi\phi}$ {are mentioned for the sake of completeness, but not} relevant for this study. 

% representing the DM flavour triplet, $\chi$ being the NP mediator and the quark-DM coupling matrix $\lambda_{ij}$. For the study we choose our DM particles to be Dirac fermions and the mediator will be a scalar. To make sure $\chi$ will constitute a DM candidate, we demand it to be a single under all SM gauge groups. As a consequence, to make the interaction term a gauge singlet, $\phi$ has to carry both the hypercharge and the colour of the right-handed up-quarks. Following the paradigm of Dark Minimal Flavour Violation, we do not constrain the quark-DM coupling $\lambda$, allowing it to be a general 3$\times$3 complex matrix, which will be a new source of both flavour and CP violation. The couplings $\lambda_{H\phi}$ and $\lambda_{\phi\phi}$ are essentially not relevant for this study. 

As in \cite{DMFVPrimer}, we parametrize the coupling matrix as
\begin{equation}
 \lambda=U_{\lambda}D_{\lambda}
\end{equation}
with a diagonal real matrix $D_{\lambda}$, and $U_{\lambda}$ consisting of three unitary matrices carrying a mixing angle and a phase each \cite{Blanke:2006xr}:
\begin{eqnarray}
D_{\lambda}&\hspace*{-1ex} = \hspace*{-1ex}&\text{diag}(D_{\lambda,11},D_{\lambda,22},D_{\lambda,33})\,,\qquad D_{\lambda,ii}>0\,, \vspace{1mm}\\
U_{\lambda}&\hspace*{-1ex} = \hspace*{-1ex}& U^{\lambda}_{23}U^{\lambda}_{13}U^{\lambda}_{12} \\[7pt]
	   &\hspace*{-1ex} = \hspace*{-1ex}& \left( \begin{matrix} 1&0&0\\ 0&c_{23}&s_{23}e^{-i\delta_{23}} \\ 0&-s_{23}e^{i\delta_{23}}&c_{23} \end{matrix} \right)
	   \left( \begin{matrix} c_{13}&0&s_{13}e^{-i\delta_{13}} \\ 0&1&0 \\ -s_{13}e^{i\delta_{13}}&0&c_{13} \end{matrix} \right)
	   \left( \begin{matrix} c_{12}&s_{12}e^{-i\delta_{12}}&0 \\ -s_{12}e^{i\delta_{12}}&c_{12}&0 \\0&0&1 \end{matrix} \right)\,.\nonumber           
\end{eqnarray}
Here $c_{ij}=\cos \theta_{ij}$ and $s_{ij}=\sin \theta_{ij}$.

This ansatz implies an unbroken $\mathbb{Z}_3$ symmetry under which only the new particles $\chi_i$ and $\phi$ are charged. This symmetry prevents the decay of any of the new particles to pure SM final states,
guaranteeing the stability of the lightest new state. For a proof of the existence of the $\mathbb{Z}_3$ symmetry, see appendix B of \cite{DMFVPrimer}, which closely follows the argument in \cite{Batell:2011tc}.

\subsection{Mass Hierarchy in the Dark Sector}

Due to the $U(3)_{\chi}$ flavour symmetry the masses of the different DM flavours are the same at the level of the tree-level Lagrangian, as in the DMFV framework the only sources of flavour violation are the SM Yukawa couplings and the new coupling matrix $\lambda$. Still an unavoidable source of DM mass splitting are effects from renormalization group running. In addition, in a complete model quantum corrections from additional heavy states can arise. Adapting the usual MFV expansion \cite{D'Ambrosio:2002ex} to the case of DMFV, we can parametrize such corrections  in terms of an expansion in the flavour violating coupling $\lambda$,
\begin{equation} \label{DMmasssplit}
m_{\chi,ij}=m_{\chi}\left(\mathds{1}+\eta\lambda^{\dagger}\lambda+\mathcal{O}(\lambda^4)\right)_{ij}=m_{\chi}\left(\mathds{1}+\eta(D_{\lambda,ii})^2 \delta_{ij}+\mathcal{O}(\lambda^4)\right)_{ij}\,,
\end{equation}
with no summation implied in the last term. Here, $\eta$ parametrizes our lack of knowledge of the full theory. Hence we will treat it as an additional parameter. The DM mass hierarchy depends both on the sign of $\eta$ and the magnitude of the couplings \lam{i}. To ensure convergence of this formula we demand $|\eta(D_{\lambda,ii})^2|<0.3$ in our analysis.

\subsection{Parameter Ranges}

To study the effects of all phenomenological constraints, we will randomly select points of the parameter space and check whether they comply with the constraining observables. To avoid double counting in the scanned parameter-space, we take the parameters of the coupling matrix $\lambda$ to lie in the following ranges:
\begin{equation}
 \delta_{ij} \in [0,2\pi), \qquad	\theta_{ij} \in [0,\frac{\pi}{4}], \qquad D_{\lambda,ii}>0.
\end{equation}
Note that this choice implies $0\le \sin\theta_{ij} \le 1/\sqrt{2}$.

In order to avoid a stable coloured particle $\phi$, we also need to make sure that
\begin{equation}
  m_{\chi} < m_{\phi}.
\end{equation}
Due to the DM mass corrections in \eqref{DMmasssplit} this could still be insufficient to ensure fermionic DM. But, as will be discussed in more detail later, the experimental constraints favour top-flavoured DM (the DM flavour coupling primarily to the SM top-quark) and $D_{\lambda,33}>D_{\lambda,11},D_{\lambda,22}$. A negative $\eta$ allows for a simultaneous compliance of those demands and in addition makes sure that all DM mass corrections will decrease the physical $m_{\chi_i}$. Hence, since $\eta$ can be treated as a free parameter, in the study we choose $\eta < 0$.

\section{Constraints from Collider Searches} 
\label{ColliderSection}

In this section we  take a look at the constraints from new physics (NP) searches at the LHC on the presented model. The obtained exclusion limits will help us to restrict the parameters of our model in a meaningful way for the further analysis. 
It has been shown in \cite{Papucci:2014iwa} that for models with a coloured $t$-channel mediator in a large fraction of the parameter space jets+$\cancel{\it{E}}_{T}$ searches are more constraining than monojet+$\cancel{\it{E}}_{T}$ searches. This observation has been confirmed for the case of bottom-flavoured DM in \cite{DMFVPrimer}, where the most stringent constraints stemmed from recasting SUSY sbottom and light squark searches. We expect the same conclusions to hold also in the present case of top-flavoured DM, and therefore restrict our study to recasting searches for top squarks and first generation squarks at the LHC.
The dominant contributions to these searches stem from the production and subsequent decay of the mediator $\phi$.

Before proceeding with our analysis, a comment is in order concerning the choice of experimental analyses taken into account. With the rapidly increasing integrated luminosity at run 2 of the LHC, the constraints on the mass scale of new particles become increasingly stringent. Already some of the early 13\,TeV analyses, using only a few fb$^{-1}$ of data, outperformed the respective searches at 8\,TeV.
In the present paper, however, we considered only the constraints from run 1 of the LHC, and disregarded the recent ones from run 2, which appeared during the completion of this work. We are aware of the fact that this approach leads to an underestimate of the constraints from the LHC. In view of other limitations of our analysis, such as considering only leading order (LO) contributions to the NP production cross-sections, and assuming the final state kinematics to be the same as in the simplified SUSY models studied by the experimental collaborations, we believe that the omission of 13\,TeV data is justifiable. We expect that while the bounds on the NP masses will shift with the inclusion of 13\,TeV data, the overall pattern of constraints will remain unaffected. A detailed study of the constraints from run 2 of the LHC is therefore left for future work.

\subsection{Production and Decay of the Mediator}

\begin{figure}[h!]
\centering

\begin{subfigure}[t]{0.4\textwidth}
\vspace{2mm}

\begin{fmffile}{NPproductionLHC}
  \centering
  \quad\quad\quad\begin{fmfgraph*}(25,20)
    \fmfbottom{i1,o1}
    \fmftop{i2,o2}
    
    \fmflabel{$q_{i}$}{i2}
    \fmflabel{$\bar{q}_{k}$}{i1}
    \fmflabel{$\phi^\dagger$}{o2}
    \fmflabel{$\phi$}{o1}
    \fmf{fermion}{v1,i1}
    \fmf{fermion}{i2,v2}
    \fmf{dashes}{o1,v1}
    \fmf{dashes}{v2,o2}
    \fmf{fermion,tension=1,label=$\chi_{j}$}{v2,v1}
  \end{fmfgraph*}
\end{fmffile}
\vspace{0.3cm}
\caption{$t$-channel $\phi$ production}
  \label{NPLHCprocesses:sfig1}
\end{subfigure}
\hspace{1cm}
\begin{subfigure}[t]{0.4\textwidth}
\vspace{2mm}

\begin{fmffile}{NPdecayLHC}
  \centering
  \quad\quad\quad\begin{fmfgraph*}(25,20)
    \fmfleft{i1}
    \fmfright{o1,o2}
    
    \fmflabel{$\phi$}{i1}
    \fmflabel{$q_i$}{o2}
    \fmflabel{$\chi_j$}{o1}
    \fmf{dashes}{v1,i1}
    \fmf{fermion}{v1,o2}
    \fmf{fermion}{o1,v1}
  \end{fmfgraph*}
\end{fmffile}
\vspace{0.3cm}
 \caption{mediator decay}
  \label{NPLHCprocesses:sfig2}
\end{subfigure}

\caption{(a) $t$-channel DM exchange diagram contributing to $\phi{\phi}^\dagger$ production. (b) Decay of the mediator $\phi$.}
\label{NPLHCprocesses}
\end{figure}
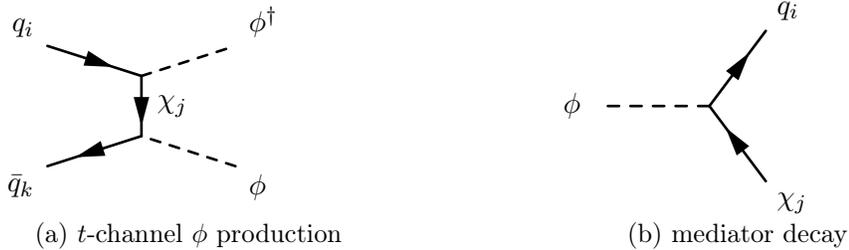

Since the mediator $\phi$ carries colour charge, it is produced via strong interaction processes at the LHC. In addition to the pure QCD process, in parts of the parameter space also the $t$-channel $\chi$ exchange diagram shown in \autoref{NPLHCprocesses:sfig1} is relevant. While the QCD production cross-section is independent of the DM mass, an $m_\chi$ dependence is introduced by the $t$-channel $\chi$ exchange diagram. Furthermore we note that since $\phi$ is charged under the new $\mathbb{Z}_3$ symmetry, it can not be singly produced --  its dominant production mode is as $\phi {\phi}^\dagger$ pairs. Also because of the $\mathbb{Z}_3$ charge, the decay of the mediator is purely governed by the interaction in \autoref{NPLHCprocesses:sfig2}. 

The relevant processes for LHC studies then are 
\begin{equation}
pp \rightarrow \phi{\phi}^\dagger \rightarrow \chi_{i} \bar\chi_{j} q_{k} \bar q_{l}\,,
\end{equation}
where $i,j,k,l$ are flavour indices.
 Depending on the quark flavours produced in the decay of the mediator $\phi$, the final state contains top quark decay products and/or light quark jets. The DM particles escape the detector and only appear as missing transverse energy. Note that for the study of collider constraints we assume the DM flavours to be degenerate. As small splittings would result in additional soft visible decay products that are difficult to search for, this approximation is justified.

\subsection{Analysis of LHC Constraints}

The final state signatures are then the same as in searches for supersymmetric squarks already conducted at the LHC.
%, see e.\,g.\ \cite{Aad:2014kra,Aad:2014wea} {\bf (Maybe different or more sources here?)}.  
The production of a stop anti-stop pair yields a $t\bar{t}+\cancel{\it{E}}_{T}$  signature, while the production and decay of squarks of the first or second generation gives jet signals with missing transverse energy. 
In addition to these experimentally well-constrained channels, also the final state $tj+\cancel{\it{E}}_{T}$ is generated, similar to supersymmetric scenarios with flavour violating squark decays, see e.\,g.\ \cite{Hurth:2009ke,Blanke:2013zxo,Agrawal:2013kha,Arana-Catania:2014ooa,Backovic:2015rwa,Blanke:2015ulx}. As no dedicated searches for this final state exist, we do not pursue it further here. 

Instead, in order to get a handle on the bounds from LHC on our model, we pick two of the most constraining run 1 squark searches by ATLAS, \cite{Aad:2014kra} for top squarks, and \cite{Aad:2014wea} for first and second generation squarks, and apply the obtained cross-section limits to our model. To this end we implement our model in {\sc FeynRules} \cite{Alloul:2013bka}, evaluate the LO signal cross section with {\sc MadGraph 5} \cite{Alwall:2014hca}, and compare the results with the respective exclusion limits presented in \cite{Aad:2014kra,Aad:2014wea}. In doing this we neglect the potentially different final state kinematics arising from the $t$-channel production process, which we deem to have a minor impact on our results.
To reduce the number of free parameters, we first set the mixing angels and phases in the coupling matrix $\lambda$ to zero. The influence of the mixing angles will later be discussed in more detail. Furthermore we assume a  coupling degeneracy \lam1=\lam2 for simplicity.

\begin{figure}
 \begin{subfigure}[t]{0.45\textwidth}
  \centering
  \includegraphics[width=\linewidth]{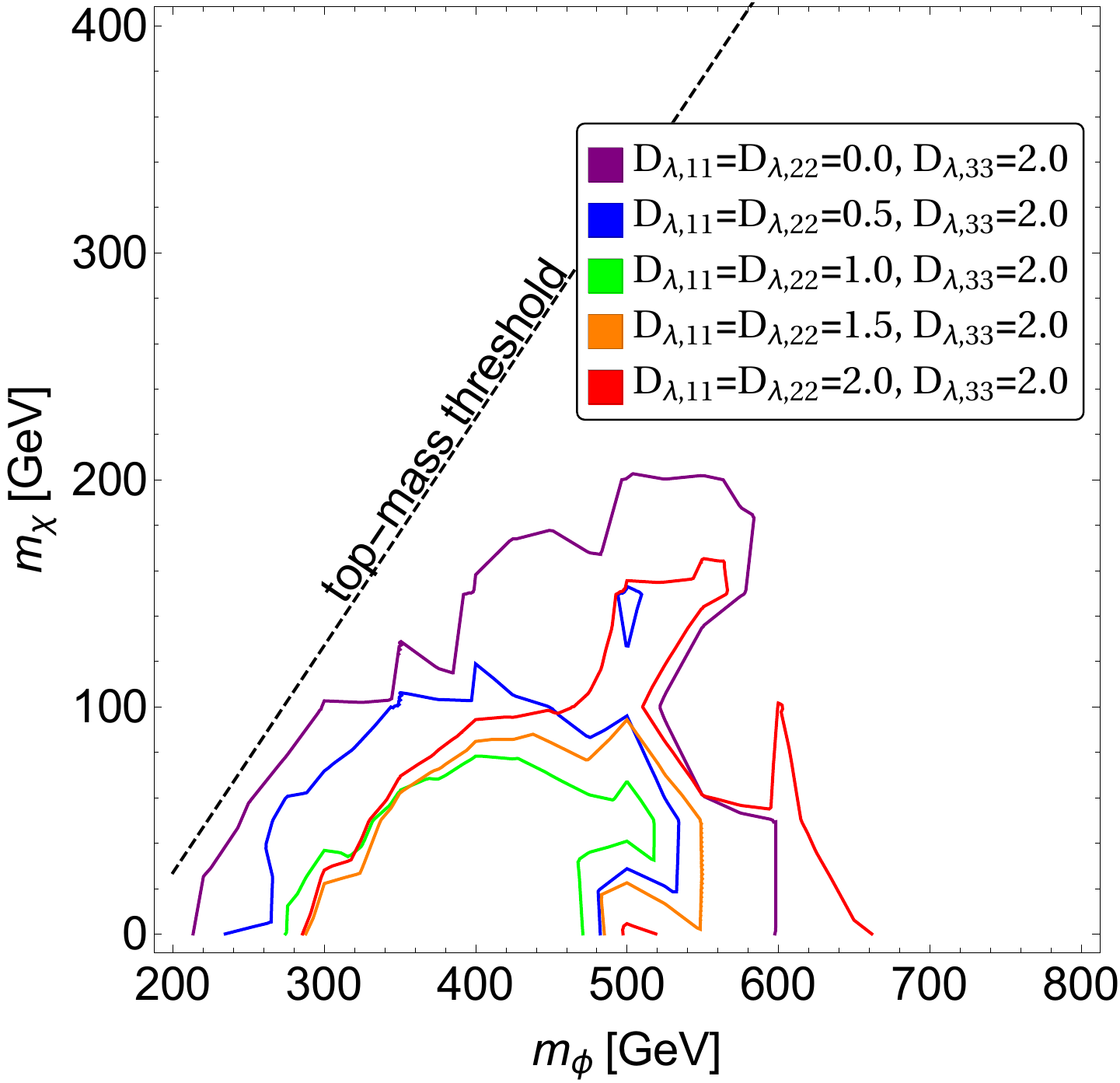}
  \caption{95\% C.L. exclusion contours for varying first and second generation couplings.}
  \label{ttbar_couplings_influence:sfig1}
 \end{subfigure} 
 \hfill
\begin{subfigure}[t]{0.5\textwidth}
  \centering
  \includegraphics[width=\linewidth]{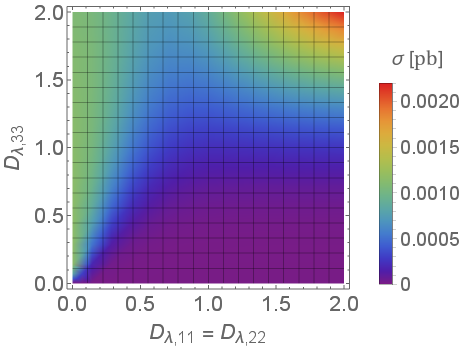}
  \caption{LO $t\bar{t}+\cancel{\it{E}}_{T}$ cross section in 8\,TeV $pp$ collisions, for \mphi=850 GeV and \mchi=50 GeV.}
  \label{ttbar_couplings_influence:sfig2}
\end{subfigure} 

\caption{Constraints from the $t\bar{t}+\cancel{\it{E}}_{T}$ final state at the 8\,TeV LHC, obtained from \cite{Aad:2014kra}.}
\label{ttbar_couplings_influence}
\end{figure}

Applying the ATLAS cross-section limits on the $t\bar{t}+\cancel{\it{E}}_{T}$ final state \cite{Aad:2014kra} to our model, we obtain the exclusion contours shown in \autoref{ttbar_couplings_influence:sfig1} for different values of \lam1=\lam2 and constant \lam3=2.0. We can see that for small \lam1=\lam2 the excluded mass range is relatively large and shrinks when those couplings are increased. The reason is the decrease of the branching ratio into the top final state, since the couplings of $\phi$ to up and charm become stronger. However we can also see that the excluded area starts to grow again, when \lam1=\lam2 grows even bigger. This effect originates in the $t$-channel production process becoming relevant. For large enough couplings this process exceeds the QCD production significantly. Due to the valence up quarks in the proton, it is in fact the value of \lam1 which governs the magnitude of this process. The effect can also be seen in \autoref{ttbar_couplings_influence:sfig2}. The $t\bar{t}+\cancel{\it{E}}_{T}$ cross-section takes the highest values for large and degenerate couplings \lam1=\lam2=\lam3, although the branching ratio into $t\bar{t}+\cancel{\it{E}}_{T}$ is only about 1/9 in this case. 

\begin{figure}
  \begin{subfigure}{0.45\textwidth}
   \centering
\includegraphics[width=\linewidth]{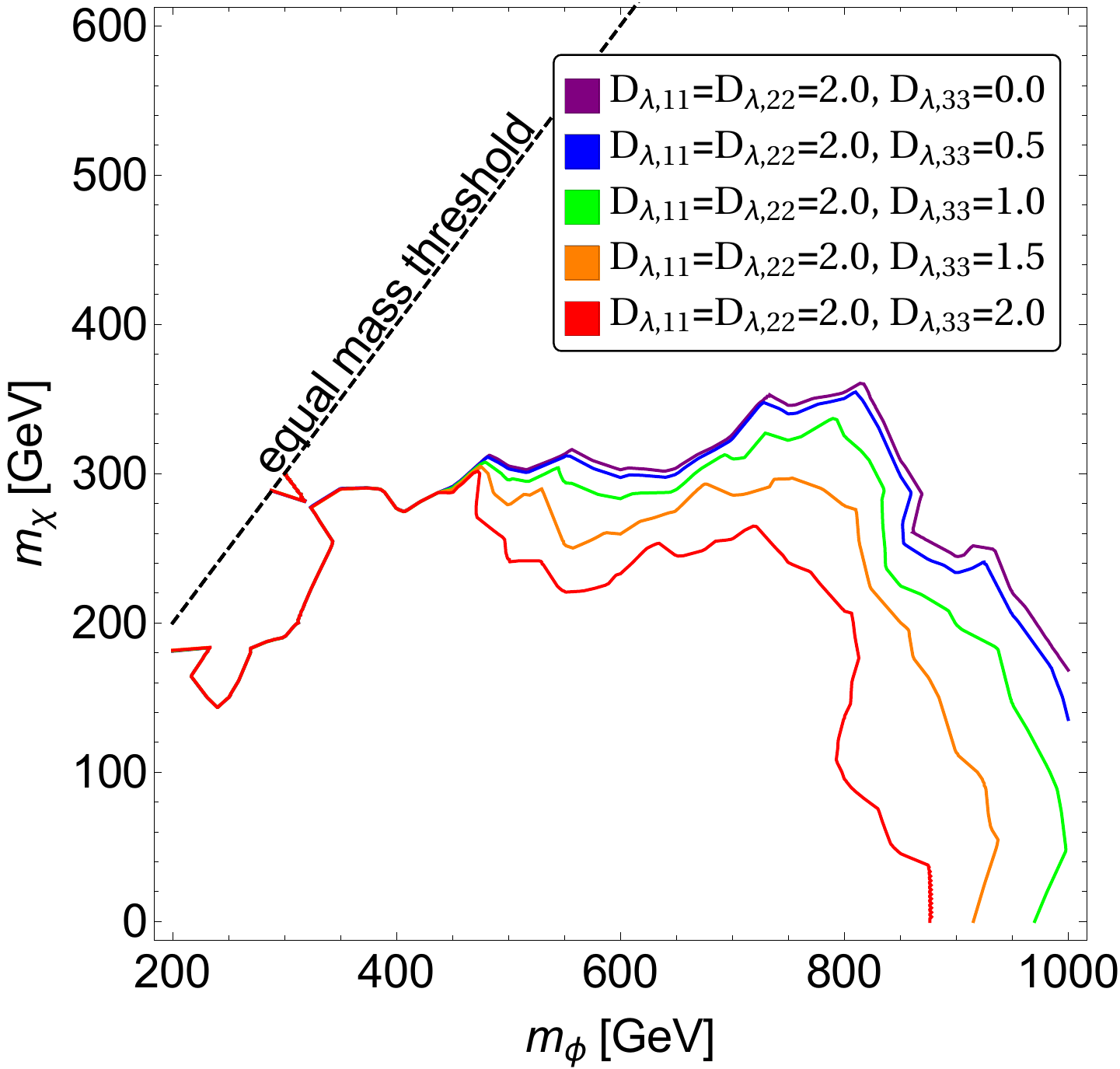}
   \caption{95\% C.L. exclusion contours with fixed \lam1=\lam2$=2.0$ and increasing \lam3. }
   \label{jj_couplings_influence:sfig1}
  \end{subfigure}
  \hfill
\begin{subfigure}{0.5\textwidth}
   \centering
\includegraphics[width=\linewidth]{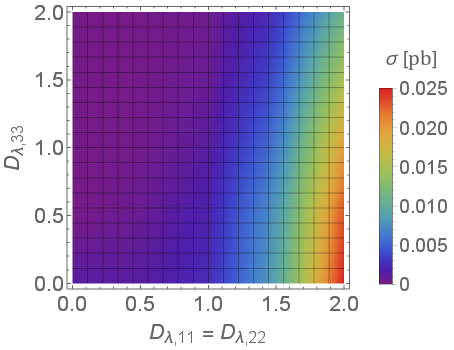}
   \caption{LO $\text{jets}+\cancel{\it{E}}_{T}$ cross section in 8\,TeV $pp$ collisions, for \mphi=850 GeV and \mchi=50 GeV.}
   \label{jj_couplings_influence:sfig2}
\end{subfigure} 

\caption{Constraints from the $\text{jets}+\cancel{\it{E}}_{T}$ final state at the 8\,TeV LHC, obtained from \cite{Aad:2014wea}.}
\label{jj_couplings_influence}
\end{figure}

Next let us take a look at the exclusion limits obtained from the ATLAS search for the $\text{jets}+\cancel{\it{E}}_{T}$ final state \cite{Aad:2014wea}. In \autoref{jj_couplings_influence:sfig1} we see the excluded areas for fixed \lam1=\lam2=2.0 with increasing \lam3. The $\phi\phi^\dagger$ production cross-section hence only depends on the masses, but the branching ratio into the jet final state decreases with increasing \lam3. We see that for $D_{\lambda,33} < D_{\lambda,11}, D_{\lambda,22}$ the constraints  nearly exclude the entire interesting parameter space. While a large mediator mass would ensure that the constraints are satisfied, in combination with the constraints from the observed relic abundance also a large DM mass would be required (see \autoref{RAsection}). Since we do not want to preclude DM masses in the region most accessible to direct DM detection experiments, we do not pursue this option.
 
The red curve, \lam1=\lam2=\lam3=2.0, still excludes a significant region of parameter space. By taking the mediator mass to be \mphi=850 GeV  and the couplings  $D_{\lambda,11} = D_{\lambda,22} < D_{\lambda,33} \le 2.0$, we can make sure that the $\text{jets}+\cancel{\it{E}}_{T}$ constraints are always fulfilled. Such a setup also allows for reasonable DM masses, as we will see later on. This choice also ensures that the $t\bar{t}+\cancel{\it{E}}_{T}$ constraints are satisfied.

\autoref{jj_couplings_influence:sfig2} illustrates the dependence of the $\text{jets}+\cancel{\it{E}}_{T}$ cross-section on the couplings \lam1=\lam2 and \lam3. Increasing values of \lam1=\lam2 increase the contribution of the $t$-channel production process. An increase in \lam3 on the other hand reduces the branching ratio of the $\text{jets}+\cancel{\it{E}}_{T}$ final state.

\subsection{Impact of Flavour Mixing Angles}

Non-zero flavour mixing angles in the coupling matrix $\lambda$ can have a significant impact on the LHC constraints discussed above. A non-zero mixing angle $\theta_{ij}$ allows the mediator to decay into a quark $q_i$ and a DM flavour $\chi_j$ not associated with this quark flavour. This decay is then governed by \lam{j}, in contrast to the flavour conserving case where the decay into the quark $q_i$ is always governed by \lam{i}. Hence, non-zero flavour mixing 
effectively decreases the influence of one \lam{i} on a quark final state in favour of another, \lam{j}. So if one of these couplings is quite small, while the other is at the upper end of the allowed parameter range, the effects could  significantly change the branching ratio into a given final state. The $t$-channel production process can be affected in a similar way. 

For the choice of parameter ranges we make at the end of this section, these effects raise no issue with the collider constraints. Since we impose \lam1,\lam2$\leq$\lam3, flavour mixing will {never be able to cause cross sections for} $\text{jets}+\cancel{\it{E}}_{T}$ {final states which are larger than in the case of} \lam1=\lam2=\lam3 {and hence the red exclusion line in} \autoref{jj_couplings_influence:sfig2}, {based on which we choose the mediator mass, remains the worst case scenario}. Due to this choice, also the constraints on $t\bar{t}\cancel{\it{E}}_{T}$ are not problematic.

\subsection{Summary of LHC Constraints} 

Summarizing the results of this section, the application of constraints from LHC searches for supersymmetric squarks yields the following information:
\begin{itemize}
 \item The $t$-channel production process plays a dominant role for large couplings \lam1=\lam2.
 \item The most stringent constraints come from searches for $\text{jets}+\cancel{\it{E}}_{T}$ final states.
 \item By appropriately restricting our parameter ranges for the studies to follow, we can ensure that the constraints from LHC searches are satisfied. We choose:
      \begin{eqnarray}
      m_\phi &=& \text{850 GeV}\,,\\
      2.0 &\geq& D_{\lambda,33} \geq D_{\lambda,11},D_{\lambda,22}\,.
      \end{eqnarray}
We note that the chosen value of $m_\phi$ might be too low to be consistent with the 13\,TeV LHC data if the DM mass is small. As we will see later, however, the DM mass is bounded from below by requiring the correct relic abundance, as well as imposing the cross-section limits from direct detection experiments.
 \item Flavour mixing angles can in general have a significant influence on the observed cross-sections. However, our choice of parameter ranges ensures that the constraints are satisfied.
\end{itemize}

\section{Flavour Constraints}
\label{Fsection}

By construction the DMFV framework allows the coupling matrix $\lambda$ to include both flavour mixing angles and CP-violating phases and therefore in general leads to significant new flavour and CP-violating effects. In the original model with DM coupling to down-type quarks \cite{DMFVPrimer}, strong constraints on the structure of the coupling matrix $\lambda$ were derived from the measured values of $K^0-\bar K^0$ and $B_{d,s}-\bar B_{d,s}$ mixing observables. These constraints are not relevant for the present model where DM couples to up-type quarks. Instead the only relevant constraints are obtained from $D^0-\bar D^0$ mixing observables.

\subsection[New Contribution to Neutral $D$ Meson Mixing]{\boldmath New Contribution to Neutral $D$ Meson Mixing}

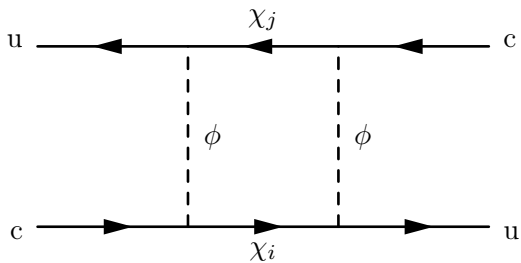
\begin{figure}[h]
\centering
\begin{fmffile}{Dmesonbox}
  \begin{fmfgraph*}(60,30)
    \fmfbottom{i1,d1,o1}
    \fmftop{i2,d2,o2}
    
    \fmflabel{c}{i1}
    \fmflabel{u}{i2}
    \fmflabel{u}{o1}
    \fmflabel{c}{o2}
    \fmf{fermion}{i1,v1}
    \fmf{fermion}{v2,o1}
    \fmf{fermion,label=$\chi_{i}$}{v1,v2}
    \fmf{fermion}{o2,v4}
    \fmf{fermion}{v3,i2}
    \fmf{fermion,label=$\chi_{j}$}{v4,v3}
    \fmf{dashes,tension=0,label=$\phi$}{v1,v3}
    \fmf{dashes,tension=0,label=$\phi$}{v2,v4}
  \end{fmfgraph*}
\end{fmffile}
\caption{Feynman diagram for the new one-loop contribution to neutral $D$ meson mixing}
\label{fig:Dmesonmix}
\end{figure}

\autoref{fig:Dmesonmix} shows the leading order NP contribution to the neutral $D$ meson mixing amplitude. Calculating the
diagram and including the appropriate symmetry factor, we find the effective Hamiltonian
\begin{equation}
\mathcal{H}_\text{eff}^{\Delta C=2,\text{new}}=\frac{1}{128\pi^{2}m_{\phi}^{2}}  \sum_{i,j}\lambda_{ui}\lambda_{ci}^{*} \lambda_{uj}\lambda_{cj}^{*}\cdot L(x_{i},x_{j})\cdot Q_{uc}^{VRR}+h.c.
\end{equation}
with the loop function
\begin{equation}\label{eq:L}
L(x_{i},x_{j})=\left(\frac{x_{i}^{2}\log(x_{i})}{(x_{i}-x_{j})(1-x_{i})^{2}}+\frac{x_{j}^{2}\log(x_{j})}{(x_{j}-x_{i})(1-x_{j})^{2}}+\frac{1}{(1-x_{i})(1-x_{j})}\right)
\end{equation}
and the effective operator 
\begin{equation}
Q_{uc}^{VRR}=\left(\bar{u}_{\alpha}\gamma_{\mu}P_{R}c_{\alpha}\right)\left(\bar{u}_{\beta}\gamma_{\nu}P_{R}c_{\beta}\right)\,,
\end{equation}
where summation over the colour indices $\alpha,\beta$ is understood. Note that throughout this section we neglect the influence of the aforementioned DM mass splittings (see \eqref{DMmasssplit}). We have checked that the splitting only causes corrections of a few percent to the loop function $L$ in \eqref{eq:L}, which are irrelevant in view of other uncertainties.

To obtain the NP contribution to the off-diagonal element of the $D^0-\bar D^0$ mass matrix, we use the expression for the hadronic matrix element:
\begin{equation}
\braket{\bar{D}^{0}|Q_{uc}^{VRR}|D^{0}}=\frac{2}{3}m_{D}^{2}f_{D}^{2}\hat{B}_{D}\,,
\end{equation}
from which we find
\begin{eqnarray}
M_{12}^{D,\text{new}} & = &\frac{1}{2m_{D}} \braket{\bar{D}^{0}|\mathcal{H}_\text{eff}^{\Delta C=2,\text{new}}|D^{0}}^{*}\\
 & = & \frac{1}{384\pi^{2}m_{\phi}^{2}} \,\eta_{D}\, m_{D}\,f_{D}^{2}\,\hat{B}_{D} \sum_{i,j}\lambda_{ui}^{*}\lambda_{ci}\lambda_{uj}^{*}\lambda_{cj}\cdot L(x_{i},x_{j})\,.\nonumber 
\end{eqnarray}
 The parameter $\eta_D$ comprises the corrections from renormalisation group running from the weak scale $\mu\sim M_W$ to the meson scale $\mu=3\,\text{GeV}$ \cite{Buras:2001ra}, where the relevant lattice calculations \cite{Aoki:2013ldr,Carrasco:2014uya} are performed. Following \cite{DMFVPrimer}, we neglect the contribution from running from the weak to the NP scale as well as differences in matching conditions between the NP scenario and the SM.

Since the new particles have significantly larger masses than the neutral $D$ mesons, the off-diagonal element of the absorptive part of the mixing amplitude, $\Gamma^D_{12}$, is unaffected by NP.

\subsection[Constraints from Neutral $D$ Meson Mixing]{\boldmath Constraints from Neutral $D$ Meson Mixing}

\begin{table}[b]
\centering
       \begin{tabular}{|c|l|}
       \hline
       \rule{0pt}{2ex}
        $m_{D^{0}}$ &
$(1864.75\pm0.15\pm0.11)\,\mbox{MeV}$\tabularnewline
       $\hat{B}_{D}$ &  $0.75\pm 0.02$\tabularnewline
       $f_{D}$ & $209.2\pm3.3\mbox{ MeV}$\tabularnewline
       $\eta_{D}$ &  $0.772$\tabularnewline
       \hline
       \rule{0pt}{3ex}
       $\tau_{D^0}$ &  $0.41\mbox{ ps}$ \tabularnewline
        \hline
        \rule{0pt}{3ex}
        $x_{12}^{D}$ & $\in\left[0.10\%,0.67\%\right]\mbox{   (95\% CL)}$\tabularnewline
       $\Phi_{12}^{D}$ & $\in\left[-5.3^\circ,4.4^\circ\right]\mbox{   (95\% CL)}$\tabularnewline
       \hline
       \end{tabular}
\caption{Parameters and experimental constraints used in the numerical
analysis \cite{Aoki:2013ldr,Carrasco:2014uya,Buras:2001ra,Aaij:2013uaa,Olive:2016xmw,Amhis:2016xyh}.}
\label{FlavourConstraintsTable}
\end{table}

Using the model-independent constraints on the $D^0-\bar D^0$ mixing amplitude \cite{Amhis:2016xyh}, as well as the numerical values for the other input parameters collected in \autoref{FlavourConstraintsTable}, we can now constrain the parameter space of our model. Recall that the CP-violating phase $\Phi_{12}^{D}$ is simply the phase of $M_{12}^{D}$, when using the convention $\arg(\Gamma_{12}^D) = 0$. In the SM, $\Phi_{12}^D$ is predicted to be of the order of $10^{-3}$ and therefore negligible. Much larger values can however be generated by the complex phases of the new coupling matrix $\lambda$. The absolute value of $M_{12}^D$ is much less precisely known in the SM, due to the dominance of long-distance contributions. In \cite{Petrov:2013usa} theoretical arguments where presented, expecting the SM contribution to 
\begin{equation}
x_{12}^{D} = \frac{2 |M_{12}^D|}{\Gamma_D} = 2 |M_{12}^D| \tau_{D^0}
\end{equation}
to be of the order 1\%. As a very conservative estimate, we assume the SM contribution to lie in the range $x_{12}^{D}\in$[-3\%,3\%].
Allowing then the values for $x_{12}^{D}$ and $\Phi_{12}^{D}$ to lie in the $95\%$ C.L. intervals in \autoref{FlavourConstraintsTable}, we can constrain the parameters of the coupling matrix $\lambda$.

Recalling the parametrization of $\lambda$ from \autoref{modelsection} and neglecting the DM mass splittings, we can simplify the sum over the DM flavours $i,j$ as
\begin{equation}
\sum_{i,j} \lambda_{ui}^{*}\lambda_{ci} \lambda_{uj}^{*}\lambda_{cj} = \left((\lambda\lambda^\dagger)_{cu}\right)^2 = \left((U_\lambda D_\lambda D_\lambda^\dagger U_\lambda^\dagger)_{cu}\right)^2 .
\end{equation}
This expression gives us a good estimate of the effects of the $D$ meson mixing constraints. Recall that $D_\lambda$ is a real diagonal matrix and $U_\lambda$ is a product of three unitary two-generation mixing matrices. We can see that in order to suppress the new contributions to $D^0-\bar D^0$ mixing, either a near degeneracy of the couplings \lam{1}$\simeq$\lam{2} or a small mixing angle \tet{12}$\simeq0$ is required. For a more detailed discussion of flavour-safe coupling scenarios in DMFV, see section 5.2 of \cite{DMFVPrimer}.

\begin{figure}[t!]
  \centering
  \includegraphics[width=.55\linewidth]{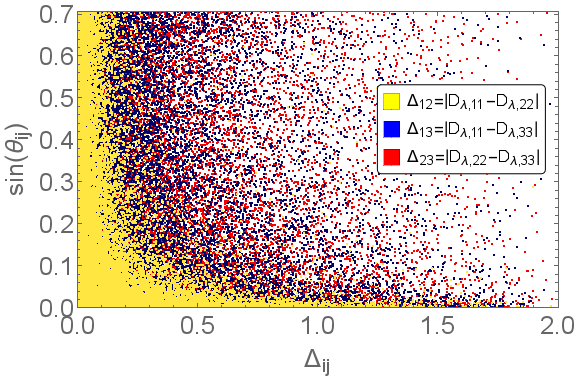}
\caption{Allowed mixing angles in dependence of the splittings between couplings \lam{i}, for DM mass \mchi=250 GeV and mediator mass \mphi=850 GeV. Different colours correspond to the different mixing angles $\theta_{ij}$ and splittings $|D_{\lambda,ii}-D_{\lambda,jj}|=\Delta_{ij}$:
$ij=12$ in yellow, $ij=13$ in blue, $ij=23$ in red.}
\label{FC_valid_area}
\end{figure}

In \autoref{FC_valid_area} we see the constraints on the flavour mixing angles $\theta_{ij}$ for a specific choice of the mediator and DM mass. We observe that, as expected, the mixing angle $\theta_{12}$ is constrained to be small, unless the couplings \lam{1} and \lam{2} are close in value.
The impact on the mixing angles $\theta_{13}$ and $\theta_{23}$, on the other hand is minor: Only if both of these mixing angles are large, a relevant NP contribution to $D^0-\bar D^0$ mixing is generated and the experimental constraints become effective. This correlation remains invisible in \autoref{FC_valid_area} since the allowed angles are shown without fixing the other parameters.
This pattern remains qualitatively the same for different values of the mediator and DM mass.

\subsection{Rare Decays}

In the original DMFV model with DM coupling to down-type quarks \cite{DMFVPrimer}, the NP effects in rare $K$ and $B$ decays have been found to be negligible. This conclusion can be transferred to our model, yielding SM-like expectations for rare decays of $D$ mesons. 

In addition, the constraints from rare flavour violating top decays are not yet stringent enough to be relevant for our model. Consequently, flavour mixing involving the third generation remains essentially unconstrained.

One possible exception is the FCNC top-quark decay $t\to q+\text{invisible}$ with $q=u,c$. Due to the large top quark mass, for a significant range of DM masses $m_\chi < m_t/2$, the decay $t \rightarrow \chi \bar{\chi} q$ is kinematically allowed and may receive a potentially large NP contribution. However, as we will see later in more detail, such low masses are excluded by a combination of relic abundance, direct detection and collider constraints.

\section{Implications of Observed Relic Abundance}
\label{RAsection}

About 80\% of all matter in the universe is dark, while only about 20\% consists of the well-studied SM particles. It is quite surprising that the order of magnitude of the matter share for SM particles and DM particles is the same. An elegant way to explain such a connection is to assume the DM abundance to be the relic of a thermal freeze-out. 
%In the early universe, at very high temperatures, SM and dark matter have been in thermal equilibrium. Due to the expansion of the universe, the temperature continued to drop, and when reaching a critical temperature $T_f$ the DM will freeze-out. Because of the NP $\mathbb{Z}_3$ symmetry, the transition of DM to SM particles is only possible via co-annihilation. Since the expansion of the universe continues, not all DM will co-annihilate and a relic abundance (RA) will remain until today. The amount of relic DM depends on the co-annihilation cross section, which depends both on the DM coupling and masses of the theory. 
For the freeze-out process to yield the observed relic abundance, theoretical considerations demand the effective cross-section  for a DM mass above 1\,GeV to be \cite{Steigman:2012nb}
\begin{equation} \label{coanpred}
\braket{\sigma v}_\text{eff}=2.2 \cdot 10^{-26}\,\mbox{cm}^{3}/\mbox{s}.
\end{equation}
We note that this constraint can be relaxed if additional stable particles contribute to the observed DM.
In our numerical analysis we require the calculated value for $\braket{\sigma v}_\text{eff}$ to
match the above value within a 10\% tolerance range. The size of this tolerance range simplifies the numerical calculation, and we checked that the results are not affected qualitatively by this choice.

In the case of flavoured DM the freeze-out process can be significantly altered by the presence of the additional dark flavours, depending on the mass splitting in the DM sector~\cite{DMFVPrimer}. Furthermore due to the large top-quark mass, the number of final states that are kinematically accessible is reduced for sufficiently small DM masses. 

In this section we first give the general expression for the DM annihilation cross section, including the relevant phase-space factors for non-negligible top quark mass. We continue with the discussion of two special cases of freeze-out scenarios and their phenomenology. 

\subsection{Annihilation of Flavoured DM}

\begin{figure}[h]
\centering
\begin{fmffile}{DMcoan}\centering

  \quad\quad\quad\begin{fmfgraph*}(40,30)
    \fmfbottom{i1,o1}
    \fmftop{i2,o2}
    
    \fmflabel{$\chi_{i}$}{i1}
    \fmflabel{$\chi_{j}$}{i2}
    \fmflabel{$q_{k}$}{o1}
    \fmflabel{$q_{l}$}{o2}
    \fmf{fermion}{v1,i1}
    \fmf{fermion}{i2,v2}
    \fmf{fermion}{o1,v1}
    \fmf{fermion}{v2,o2}
    \fmf{dashes,tension=1,label=$\phi$}{v1,v2}
  \end{fmfgraph*}
\end{fmffile}
\caption{DM annihilation process at tree-level}
\label{DMcoan}
\end{figure}
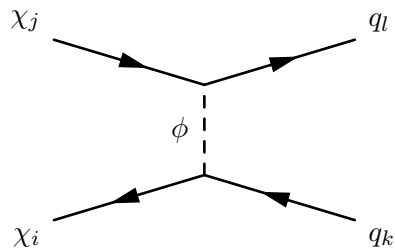

Since our model includes three DM flavours, DM annihilation can proceed via several tree level processes with different initial and final state flavours, as shown in \autoref{DMcoan}.
Depending on the flavours present at the time of freeze-out, we need to average over the possible processes to obtain the correct annihilation cross-section. First we assume that all flavours are present and the DM mass is larger than the top quark mass, so that all quark final states are kinematically allowed. Combining the procedure in \cite{Wells:1994qy} with the previous results from \cite{DMFVPrimer} (which neglected the phase-space factors due to negligible quark masses in the down-sector), we find the overall averaged annihilation cross-section 
\begin{equation} \label{coancs}
\braket{\sigma v}_\text{eff}=\frac{1}{18}\cdot\frac{3}{32\pi}\cdot\frac{1}{4}\sum_{i,j=1}^3\sum_{k,l=u,c,t}|\lambda_{ki}|^2|\lambda_{lj}|^2\,\frac{\sqrt{\left(4m_{\chi}^{2}-(m_{k}-m_{l})^{2}\right)\left(4m_{\chi}^{2}-(m_{k}+m_{l})^{2}\right)}}{\left(m_{\phi}^{2}+m_{\chi}^{2}-\frac{m_{k}^{2}}{2}-\frac{m_{l}^{2}}{2}\right)^{2}}
\end{equation}
where $m_{k,l}$ are the masses of the final state quarks. Note that this formula includes a factor of $1/2$ stemming from the conversion to an effective cross-section for a Dirac fermion \cite{Griest:1990kh, Servant:2002aq}, as well as a flavour averaging factor of $1/9$. The DM flavours present at the time of freeze-out need to have nearly degenerate masses, hence it is safe to set $m_{\chi_i}=m_{\chi_j}=m_\chi$.

\subsection{Possible Freeze-Out Scenarios}

DM freeze-out occurs when the temperature drops below a critical value $T_f\approx {m_\chi}/{20}$. As discussed in \autoref{modelsection}, in DMFV the masses of the dark flavours $\chi_i$ are split by the non-universality of the coupling matrix $\lambda$:
\begin{equation}
m_{ij}=m_{\chi}\left(\mathds{1}+\eta\lambda^{\dagger}\lambda+\mathcal{O}(\lambda^4)\right)_{ij}=m_{\chi}\left(\mathds{1}+\eta(D_{\lambda,ii})^2+\mathcal{O}(\lambda^4)\right)_{ij}. %\tag{\ref{DMmasssplit}}
\end{equation}
If the splitting is negligible compared to $T_f$, all flavours are present at the time of freeze-out and \eqref{coancs} gives the correct value for the annihilation cross-section. Although all flavours contribute to the freeze-out, the heavier flavours eventually decay and the presently observed DM consists of the lightest flavour only. If on the other hand the lightest flavour(s) is/are split significantly, the heavier flavours  have decayed by the time the DM freeze-out sets in. In that case \eqref{coancs} has to be modified accordingly: the first sum then runs only over the DM flavours present at freeze-out. As already mentioned, we focus on the phenomenologically preferred case of top-flavoured DM. In \autoref{ColliderSection} we found that the LHC constraints prefer \lam3 to be the largest DM coupling. Choosing a negative $\eta$ hence ensures the top-flavour to be the DM candidate.

In our analysis we study two benchmark cases:
\begin{itemize}
 \item In the \textbf{quasi-degenerate freeze-out (QDF)} scenario we assume all flavours to be present at the time of DM freeze-out. To ensure this we demand the mass splitting to be below 1\,\%. For simplicity we fix $\eta=-0.01$, which is the smallest justifiable magnitude. A different choice of $\eta$ changes our findings quantitatively but not qualitatively.
 \item In the \textbf{single flavour freeze-out (SFF)} scenario we focus on the case of the top-flavour being split significantly from the others and hence being the only flavour present at the time of freeze-out. We demand a mass splitting of at least 10\% for this scenario to happen. In this case we set $\eta=-0.075$ which, for our choice $D_{\lambda,ii}\le 2.0$, is the maximum value consistent with DM mass corrections of at most 30\,\%. 
\end{itemize}
In addition to this discrimination of possible freeze-out scenarios, we also need to consider the case $m_\chi < m_{t} $. If the DM mass drops below the top mass threshold, annihilation into $t\bar t$ pairs is kinematically excluded. If $m_\chi < m_{t}/2 $ also single-top final states become inaccessible. Therefore \eqref{coancs} has to be modified accordingly, affecting the constraints on the parameters of $\lambda$.

\subsection{Phenomenological Analysis of Freeze-Out Scenarios}

\subsubsection{Quasi-Degenerate Freeze-Out}

To understand the effect of the relic abundance constraints on the parameters of our model, let us first study equation \eqref{coancs} in more detail. Neglecting the phase-space factors, i.\,e.\ taking the limit $m_\chi \gg m_t$, the formula simplifies to
\begin{equation}\label{coancs-simp}
\braket{\sigma v}_\text{eff}=\frac{1}{18}\cdot\frac{3}{32\pi}\sum_{i,j=1,2,3}\frac{D_{\lambda,ii}^{2}D_{\lambda,jj}^{2}\cdot m_{\chi}^{2}}{\left(m_{\phi}^{2}+m_{\chi}^{2}\right)^{2}}.
\end{equation}

For a given pair of \mchi\ and \mphi, \eqref{coancs-simp} reduces to a spherical constraint on the couplings, i.\,e.
\begin{equation}
D_{\lambda,11}^2+D_{\lambda,22}^2+D_{\lambda,33}^2 =\text{const.}
\end{equation}
 Since all couplings \lam{i} can be taken positive without loss of generality, we are in fact limited to $1/8$ of the surface of a sphere. The constraint \lam{i}\,$<2.0$ then cuts out part of the remaining shell.  Finally reinserting the phase-space factors and therefore the angular dependence deforms the sphere. What remains as allowed coupling region is hence part of this deformed sphere. Furthermore the mass splitting conditions for QDF establish a lower bound for \lam1, \lam2 in dependence of \lam3. This dependence further shapes the allowed parameter space.

\begin{figure}[t!]
  \centering
  \includegraphics[width=.6\textwidth]{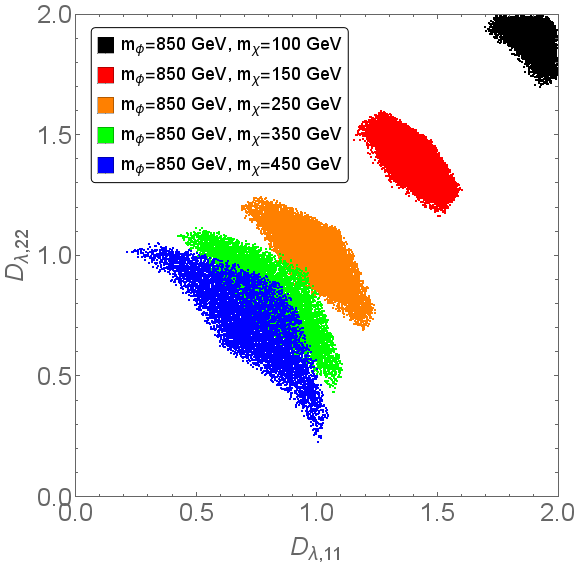}
  \caption{Regions of parameter space compatible with the relic abundance constraint in the QDF scenario, for different DM masses.\label{RA_valid_area:sfig1}}
\end{figure}

\autoref{RA_valid_area:sfig1} shows the allowed coupling range for the QDF scenario with different DM masses in the \lam1-\lam2 plane. We can clearly see the $m_\chi$ dependence of the relic abundance constraint. The smaller the DM mass gets, the larger the couplings have to be to still reach the required annihilation cross-section. This shift is more significant when \mchi\ drops below the top-mass threshold and then especially half the top-mass threshold. Since fewer and again fewer final states remain accessible below the respective thresholds, the total number of terms contributing to the cross-section decreases and the coupling parameters have to be even larger to compensate that. We can also see that the relic abundance constraints, together with the LHC limit of 2.0 for the couplings \lam{i}, establishes a lower bound on the DM mass depending on the value of the mediator mass. 
%The larger the mediator mass, the larger the combination of DM mass and coupling needs to be. Hence, below a certain threshold value the DM mass will demand couplings beyond 2.0 and hence be excluded. 
The influence of the flavour mixing angles $\theta_{ij}$, on the other hand, is insignificant for the QDF scenario. 
%Since a large number of parameters is involved in the formula, no specific angle gets seriously constrained in general.

\subsubsection{Single Flavour Freeze-Out}
\label{RAsection:SFF}

As far as flavour mixing angles are concerned, the single flavour freeze-out (SFF) scenario is an entirely different story. Since only the top-flavour $\chi_t$ is present at the time of freeze-out, only one initial state remains in \eqref{coancs} -- hence only terms involving \lam3, \tet{13} and \tet{23} contribute and the averaging factor drops out. As a result these few remaining parameters get more stringently constrained. 

\begin{figure}[t]
  \centering
  \includegraphics[width=0.6\textwidth]{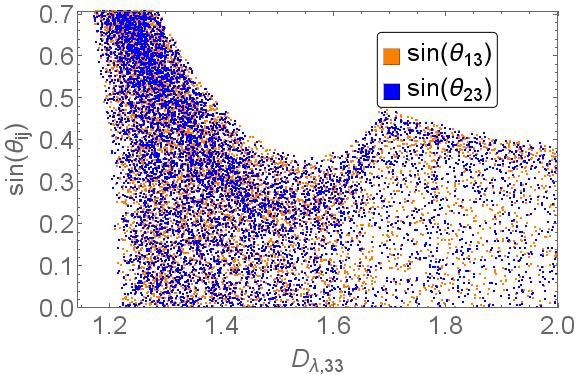}
  \caption{Allowed ranges for the mixing angles in dependence on \lam3, for SFF with \mphi=850 GeV and \mchi=220 GeV. Different colours correspond to the different mixing angles $\theta_{ij}$: $ij=13$ in orange, $ij=23$ in blue.}
  \label{RA_valid_area:sfig2}
\end{figure}

\autoref{RA_valid_area:sfig2} shows the mixing angles $\theta_{13}$ and $\theta_{23}$ as functions of  \lam3 for a mediator mass of 850\,GeV and a Lagrangian  DM mass parameter of $m_\chi = 220\,\text{GeV}$. We observe that for the smallest allowed \lam3 the mixing angles need to be maximal, in order to push the cross-section into the tolerance interval of the constraints. With increasing coupling \lam3, smaller angles are allowed as well and instead an upper bound arises. This upper bound becomes weaker when approaching the threshold value \lam3\,$\simeq 1.7$. At this point, the physical DM mass $m_{\chi_t}=m_\chi \left(1+\eta D_{\lambda,33}^2\right)$ drops below the top quark mass, so that the annihilation channel into $t\bar t$ pairs becomes inaccessible. Consequently, larger flavour mixing is again required in order to enhance the DM annihilation cross-section into light flavours.

For the same reasons as in the QDF scenario we find a lower bound on the DM mass, depending on the mediator mass. But in addition, in the SFF scenario for a given value of $\eta$ we also find an upper bound on the DM mass. The origin of this effect is a combination of the relic abundance constraints and the splitting condition for SFF. For a given mediator mass the combination of DM mass and the coupling parameters has to be in a certain interval. With increasing DM mass, the coupling hence has to take smaller values. At the same time the SFF condition demands \lam3 to be large enough to ensure a splitting of at least 10\%, hence establishing a lower bound on \lam3 depending on $\eta$, which translates into an upper bound on the DM mass. Larger DM masses require larger values of
  $m_\phi$ and/or $\eta$.

\section{Constraints from Direct Detection Experiments}
\label{DDsection}

In this section we discuss the constraints from direct detection experiments. The currently most stringent cross-section limits are provided by the LUX collaboration \cite{Akerib:2016vxi}, which we will use in our analysis. Note that a comparable sensitivity has also been reached by PandaX-II \cite{Tan:2016zwf}. We first discuss the relevant WIMP-nucleon interactions and study their interplay, resulting in an effective suppression of WIMP-xenon scattering. The stringent constraints on the WIMP-nucleon scattering cross-section will help us to make a case for top-flavoured DM, instead of the up or charm case. 

We then turn our attention to the relative abundance of stable and quasi-stable xenon isotopes in natural xenon, which is used in experiments such as LUX. The simultaneous suppression of the respective WIMP-nucleus cross sections, necessary to keep the combined effective cross section for natural xenon in check, proves to be hard to achieve. Especially in light of future experiments, such as XENON1T \cite{Diglio:2016stt}, XENONnT \cite{Diglio:2016stt}, LUX-ZEPLIN (LZ) \cite{Akerib:2015cja} and DARWIN \cite{Aalbers:2016jon}, this will result in strong constraints on the model studied in this paper.

\subsection{Relevant WIMP-Nucleon Interactions}

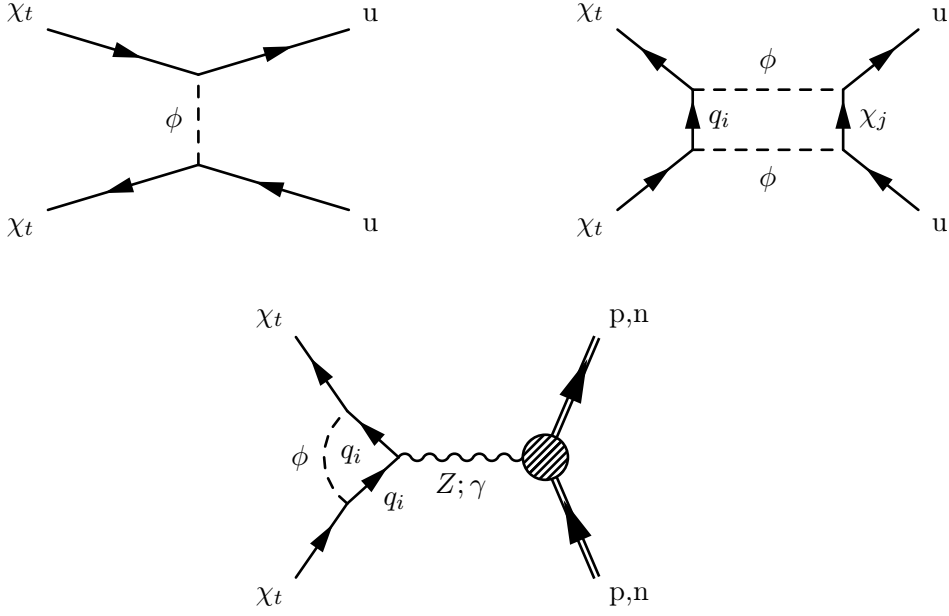
\begin{figure}[h]

\begin{fmffile}{DDTreelevel}
\centering

  \quad\quad\quad\begin{fmfgraph*}(40,30)
    \fmfbottom{i1,o1}
    \fmftop{i2,o2}
    
    \fmflabel{$\chi_{t}$}{i1}
    \fmflabel{$\chi_{t}$}{i2}
    \fmflabel{u}{o1}
    \fmflabel{u}{o2}
    \fmf{fermion}{v1,i1}
    \fmf{fermion}{i2,v2}
    \fmf{fermion}{o1,v1}
    \fmf{fermion}{v2,o2}
    \fmf{dashes,tension=1,label=$\phi$}{v1,v2}
  \end{fmfgraph*}
\end{fmffile}
\quad\quad\quad\quad\quad\quad\quad\quad
\begin{fmffile}{DDBox}
  \begin{fmfgraph*}(40,30)
    \fmfbottom{i1,d1,o1}
    \fmftop{i2,d2,o2}
    
    \fmflabel{$\chi_{t}$}{i1}
    \fmflabel{$\chi_{t}$}{i2}
    \fmflabel{u}{o1}
    \fmflabel{u}{o2}
    \fmf{fermion}{i1,v1}
    \fmf{fermion}{v2,i2}
    \fmf{fermion,label=$q_{i}$}{v1,v2}
    \fmf{fermion}{v4,o2}
    \fmf{fermion}{o1,v3}
    \fmf{fermion,label=$\chi_{j}$}{v3,v4}
    \fmf{dashes,tension=0.5,label=$\phi$}{v1,v3}
    \fmf{dashes,tension=0.5,label=$\phi$}{v2,v4}
  \end{fmfgraph*}
\end{fmffile}

\quad

\quad\quad\quad\quad\quad\quad\quad\quad

\quad\quad\quad\quad\quad\quad\quad\quad
\hspace{1cm}
\begin{fmffile}{DDpenguins}
  \begin{fmfgraph*}(40,40)
    \fmfbottom{i1,o1}
    \fmftop{i2,o2}
    
    \fmflabel{$\chi_{t}$}{i1}
    \fmflabel{$\chi_{t}$}{i2}
    \fmflabel{p,n}{o1}
    \fmflabel{p,n}{o2}
    \fmf{fermion}{i1,v1}
    \fmf{fermion}{v2,i2}
    \fmf{fermion,label=$q_{i}$}{v1,v3,v2}
    \fmf{double_arrow}{o1,v4}
    \fmf{double_arrow}{v4,o2}
    \fmfblob{0.15w}{v4}
    \fmf{dashes,left=0.5,tension=0.3,label=$\phi$}{v1,v2}
    \fmf{photon,tension=0.7,label=$Z;\gamma$}{v3,v4}
  \end{fmfgraph*}
\end{fmffile}
\caption{Feynman diagrams for relevant WIMP-nucleon interactions.}
\label{WIMP-nuleon scattering}
\end{figure}

The relevant WIMP-nucleon scattering processes are depicted in \autoref{WIMP-nuleon scattering}. The leading process is the tree-level interaction, followed by contributions from box diagrams, as well as photon- and $Z$-penguin diagrams. In contrast to the bottom-flavoured DM model studied in \cite{DMFVPrimer}, the $Z$-penguin contribution is no longer negligible, due to the large top quark mass.

The spin-independent WIMP-nucleon cross-section, for a nucleus with mass number $A$ and atomic number $Z$, can be written as
\begin{equation} \label{WIMPnucleonCS}
 \sigma_{n}^{SI} = \frac{\mu_n^2}{\pi A^2}|Z f_p + (A-Z)f_n|^2\,,
\end{equation}
with the reduced mass $\mu_n$ of the WIMP-nucleon system, and $f_p$ and $f_n$ parametrizing the DM couplings to the proton and neutron, respectively. 

Assuming, as done throughout our analysis, that the observed DM consists solely of the top-flavour $\chi_t$, and using the results of  \cite{Kumar:2013hfa,DMFVPrimer}, we find the following relevant contributions to $f_p$ and $f_n$:
\begin{itemize}
 \item At tree-level, DM couples to the up quarks in the nuclei via an $s$-channel $\phi$ exchange. We obtain: 
  \begin{equation}
  f^\text{tree}_p=2f^\text{tree}_n=\frac{|\lambda_{ut}|^2}{4m^2_\phi}. 
 \end{equation}
 \item The box-diagram contribution reads
 \begin{equation}
  f^\text{box}_p=2f^\text{box}_n=\sum_{i,j}\frac{|\lambda_{ui}|^2|\lambda_{jt}|^2}{32\pi^2m^2_\phi}L\left(\frac{m^2_{q_i}}{m^2_\phi},\frac{m^2_{\chi_j}}{m^2_\phi}\right)\,,
 \end{equation}
with the loop function $L$ given in equation \eqref{eq:L}.
 \item As the photon couples to the electric charge of the nucleon, the photon penguin diagram only contributes to $f_p$:
 \begin{equation}
 f_p^\text{photon}=-\sum_i\frac{|\lambda_{it}|^2e^2}{48\pi^2m^2_\phi}\left[\frac{3}{2}+\log\left(\frac{m^2_{q_i}}{m^2_\phi}\right)\right]
 \end{equation}
 \item Last but not least, the contributions from  the $Z$ penguin diagram are:
\begin{eqnarray}
 f_p^{Z}&=& - \frac{3|\lambda_{tt}|^2e^2\left(\frac{1}{2}-2 \sin^2\theta_W \right)}{32\pi^2 \sin^2\theta_W \cos^2\theta_W m_Z^2}\frac{m^2_{t}}{m^2_\phi}\left[1+\log\left(\frac{m^2_{t}}{m^2_\phi}\right)\right]\,,\\
 f_n^{Z}&=& - \frac{3|\lambda_{tt}|^2e^2\left(-\frac{1}{2}\right)}{32\pi^2 \sin^2\theta_W \cos^2\theta_W m_Z^2}\frac{m^2_{t}}{m^2_\phi}\left[1+\log\left(\frac{m^2_{t}}{m^2_\phi}\right)\right]\,,
\end{eqnarray}
with $\theta_W$ being the weak mixing angle. It is apparent that the latter contributions are only relevant for a top quark in the loop, since otherwise the factor of ${m^2_{q}}/{m^2_\phi}$ makes the term negligible.

\end{itemize}

\subsection{Direct Detection Constraints} \label{DD_interplay}

The sum of all contributions to the WIMP-nucleon scattering in \eqref{WIMPnucleonCS} has to give a cross-section below the bound provided by the LUX experiment \cite{Akerib:2016vxi}. This constraint is quite stringent for WIMP DM. A destructive interference of the various contributions, as found in \cite{DMFVPrimer}, is therefore needed. Taking a closer look at the individual contributions we find that only the $Z$-penguin coupling to the neutron is negative, while all other terms in \eqref{WIMPnucleonCS} are positive. Since the tree-level diagram in general yields the largest positive contribution, its cancellation with the $Z$-penguin coupling to neutrons is necessary.
 This required cancellation is  expected to have a major influence on the shape of the allowed parameter space, when imposing the direct detection bounds.

\begin{figure}[t!]
 \begin{subfigure}[t]{0.475\textwidth}
  \centering
  \includegraphics[width=\linewidth]{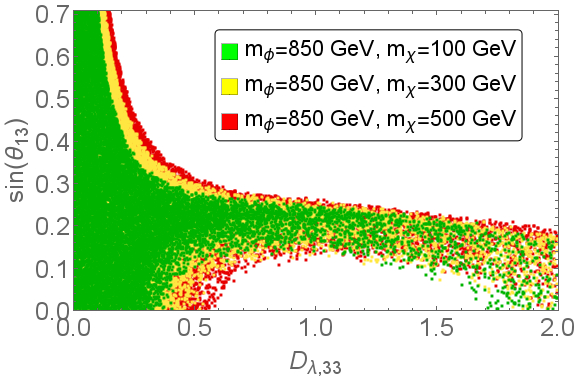}
  \caption{Mixing angle $\theta_{13}$ as function of \lam{3} for various values of $m_\chi$ and fixed $m_\phi$.}
  \label{DD_valid_area:sfig1}
 \end{subfigure} 
 \hspace{0.3cm}
\begin{subfigure}[t]{0.475\textwidth}
  \centering
  \includegraphics[width=\linewidth]{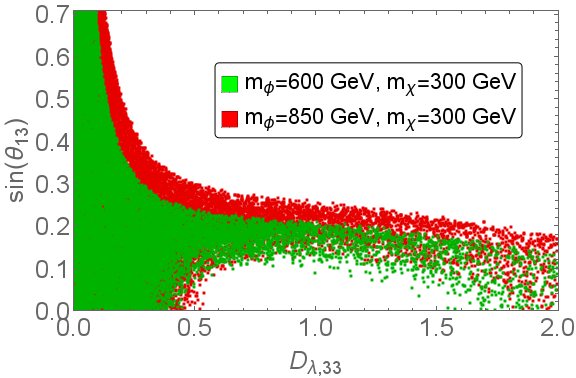}
  \caption{Mixing angle $\theta_{13}$ as function of \lam{3} for various values of $m_\phi$ and fixed $m_\chi$.}
  \label{DD_valid_area:sfig2}
\end{subfigure} \quad

\caption{Constraints on the mixing angle $\theta_{13}$ from the LUX data.}
\label{DD_valid_area}
\end{figure}

\autoref{DD_valid_area} shows the allowed region of parameter space for several values of the DM  and mediator masses, with \tet{13} plotted against \lam3. We see that for small couplings \lam3 the constraints have no impact on the allowed values of \tet{13}. This comes as no surprise, since every single contribution to the WIMP-nucleon scattering is proportional to $D^2_{\lambda,33}$, so the cross-section \eqref{WIMPnucleonCS} is proportional to $D^4_{\lambda,33}$. Sufficiently small couplings \lam3, on their own, can thus ensure that the predicted cross-section is below the LUX bound. For larger \lam3, this suppression is no longer sufficient, so that the aforementioned destructive interference between the tree-level and $Z$-penguin contributions is required. The tree-level contribution is proportional to $D_{\lambda,33}^2 \sin^2\theta_{13}$, while the $Z$-penguin is proportional to  $D_{\lambda,33}^2 \cos^2\theta_{13} \cos^2\theta_{23}$. The latter has to cancel the tree-level (and in most cases also other sub-leading positive) contributions. The sum of those two major terms has its minimum at $\sin\theta_{13}\simeq 0.2$, as can be seen from \autoref{DD_valid_area}. Due to the influence of several other parameters, such as \tet{23}, we observe an allowed interval around the value of $\sin\theta_{13}=0.2$. 
For values of \lam3 larger than unity, the box diagram contribution, being proportional to $D_{\lambda,33}^4$, becomes competitive. Consequently the allowed range for \tet{13} is driven to lower values.

The necessity of this destructive interference also helps to motivate the case for top-flavoured DM. Since the nuclei consist of  first generation quarks, choosing a DM associated with the second or third generation helps to suppress the tree-level contribution by a small flavour mixing angle. Furthermore the crucial negative contribution from the $Z$-penguin is only significant for a top-quark in the loop. Top-flavoured DM  is hence favoured by the absence of signal in direct detection experiments.

As the tree level and $Z$-penguin diagrams yield different WIMP-nucleon copling strengths for protons and neutrons, their cancellation requires a fixed proton-to-neutron ratio in the detector material. Top-flavoured DM in DMFV therefore constitutes a concrete example of xenophobic DM \cite{Feng:2013fyw}. A similar situation was encountered in the case of bottom-flavoured DM in DMFV, where the destructive interference arose between the box and photon penguin contributions \cite{DMFVPrimer}.

\subsection{Natural Xenon and its Isotopes} 
\label{XenonIsotopeSection}

\begin{table}[h]
\centering
       \begin{tabular}{|c|c|r|}
       \hline
       \rule{0pt}{2ex}
        {\bf isotope} & {\bf half-life} & {\bf\boldmath  abundance $\rho$} \tabularnewline
        \hline
       \rule{0pt}{3ex}
        $^{124}$Xe & stable & $ 0.095\% $ \tabularnewline
        $^{126}$Xe & stable & $ 0.089\% $ \tabularnewline
        $^{128}$Xe & stable & $ 1.910\% $ \tabularnewline
        $^{129}$Xe & stable & $ 26.401\% $ \tabularnewline
        $^{130}$Xe & stable & $ 4.071\% $ \tabularnewline
        $^{131}$Xe & stable & $ 21.232\% $ \tabularnewline
        $^{132}$Xe & stable & $ 26.909\% $ \tabularnewline
        $^{134}$Xe & stable & $ 10.436\% $ \tabularnewline
        $^{136}$Xe & $2.165 \times 10^{21} $ y & $ 8.857\% $ \tabularnewline
       \hline
       \end{tabular}
\caption{Xenon isotopes with respective half-life and natural abundance.}
\label{XenonIsotopesTable}
\end{table}

The LUX experiment uses natural xenon as detector material. Natural xenon consists of nine stable and quasi-stable isotopes, see \autoref{XenonIsotopesTable}. In our discussion so far we have neglected this fact and merely used an average mass number. To take into account the various isotopes, it is necessary to calculate a combined effective WIMP-nucleon cross-section for natural xenon $\sigma_{n,\text{nat-Xe}}^{SI}$, by weighting the respective cross-sections $\sigma_{n,i}^{SI}$ of the isotopes $i$ with their relative abundance $\rho_i$:
\begin{equation} \label{WIMPnucleonCSnat}
 \sigma_{n,\text{nat-Xe}}^{SI} = \sum_{i=1}^{9} \rho_i \cdot \sigma_{n,i}^{SI} = \sum_{i=1}^{9} \rho_i \cdot \frac{\mu_n^2}{\pi A_i^2}|Z f_p + (A_i-Z)f_n|^2\,.
\end{equation}

As the cancellation between positive and negative contributions to the WIMP-nucleus scattering cross-section depends on the relative weight of $f_p$ and $f_n$ and hence on $A_i$, the cancellation will require different parameter ranges for each isotope. Since several of the xenon isotope have a significant natural abundance, the combined cross-section can only remain below the bounds from direct detection experiments if every single contribution from the different isotopes is kept sufficiently small. For this to happen simultaneously, for an increasing \lam3 an increasing amount of fine-tuning in the other parameters is necessary. 

\begin{figure}[t!]
  \centering
  \includegraphics[width=.6\linewidth]{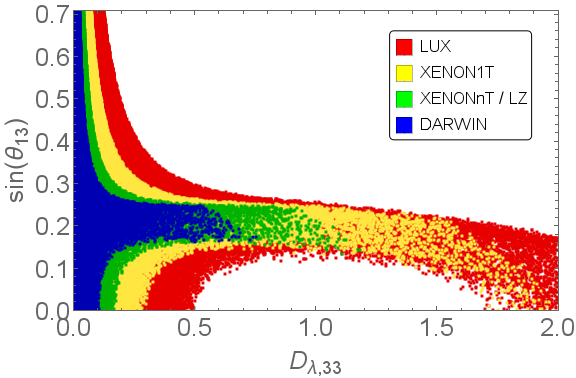}
\caption{Allowed regions of parameter space, imposing the (expected) cross-section limits from present and future direct detection experiments, for DM mass \mchi=250 GeV and mediator mass \mphi=850 GeV.}
\label{DDExpMultiBounds}
\end{figure}

We see in \autoref{DDExpMultiBounds} that the current bounds from LUX are not stringent enough to exclude any values of \lam3\,$< 2.0$. However, future direct detection experiments such as XENON1T \cite{Diglio:2016stt}, XENONnT \cite{Diglio:2016stt}, LUX-ZEPLIN (LZ) \cite{Akerib:2015cja} and DARWIN \cite{Aalbers:2016jon} will push the cross-section limit further down and hence make a sufficient supression of the cross-section $\sigma_{n,\text{nat-Xe}}^{SI}$ more difficult to accomplish. As we can see in \autoref{DDExpMultiBounds}, already XENON1T will make the cancellation impossible for large values of \lam3, and XENONnT/LZ and DARWIN will push the upper bound on \lam3 to significantly smaller values. In the next section we will see that these findings have drastic consequences in combination with the constraints from the observed relic abundance.

\section{Combined Analysis of Flavour and Dark Matter Constraints}
\label{combinedsection}

After studying the impact of flavour, relic abundance and direct detection constraints one by one, we now turn to their combined analysis. We will see that their interplay limits the allowed region of parameter space of our model in a more stringent way.

\subsection{Phenomenological Analysis}

\begin{figure}
 \begin{subfigure}{0.5\textwidth}
  \centering
  \includegraphics[width=.95\linewidth]{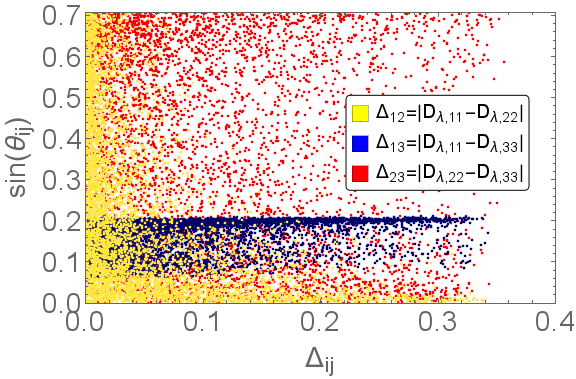}
  \caption{QDF, $m_\chi=150$ GeV}
  \label{Com-deg:sfig1}
 \end{subfigure}
\begin{subfigure}{0.5\textwidth}
  \centering
  \includegraphics[width=.95\linewidth]{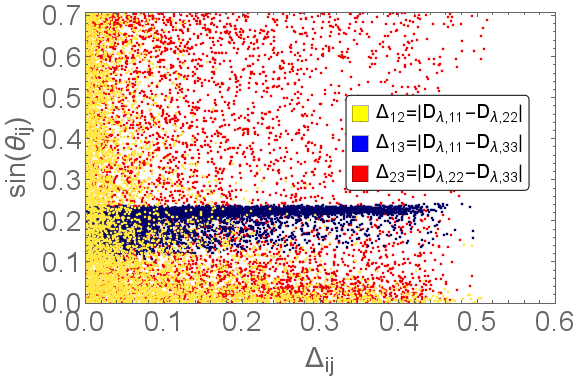}
  \caption{QDF, $m_\chi=250$ GeV}
  \label{Com-deg:sfig2}
\end{subfigure}\vspace{3mm}

\begin{subfigure}{0.5\textwidth}
  \centering
  \includegraphics[width=.95\linewidth]{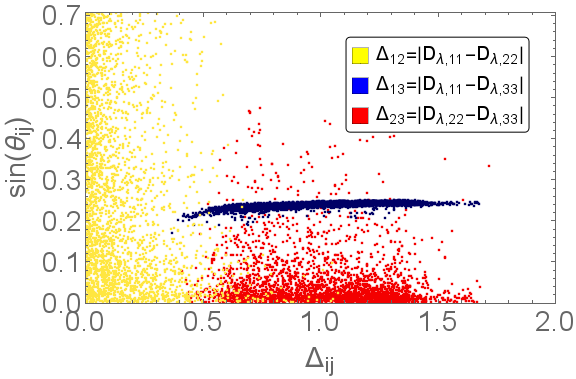}
  \caption{SFF, $m_\chi=225$ GeV}
  \label{Com-deg:sfig3}
\end{subfigure}
\begin{subfigure}{0.5\textwidth}
  \centering
  \includegraphics[width=.95\linewidth]{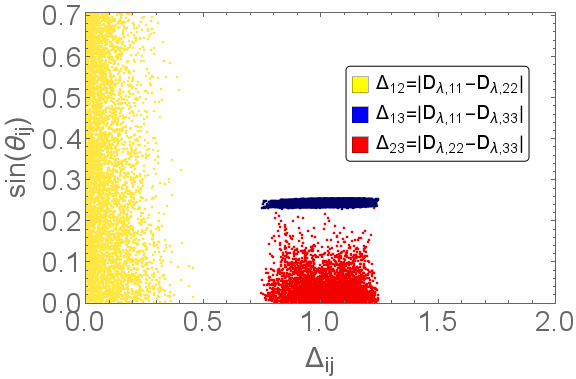}
  \caption{SFF, $m_\chi=250$ GeV}
  \label{Com-deg:sfig4}
\end{subfigure}
\caption{Viable regions of parameter space, imposing all constraints, for different freeze-out scenarios and DM masses, with $m_\phi=850$ GeV. Different colours correspond to the different mixing angles $\theta_{ij}$ and splittings $|D_{\lambda,ii}-D_{\lambda,jj}|=\Delta_{ij}$:
$ij=12$ in yellow, $ij=13$ in blue, $ij=23$ in red.}
\label{Com-deg}
\end{figure}

 \autoref{Com-deg} shows the effects of combining the various constraints on the allowed ranges for the mixing angles $\theta_{ij}$, depending on the splitting between the respective couplings.
 In all four diagrams we recover the $D^0-\bar D^0$ mixing constraint on \tet{12} -- only for a small splitting between \lam{1} and \lam{2} large values for this mixing angle  are allowed.

We also observe a strong impact on the allowed range of \tet{13}. Since a small \lam{3} is excluded by the relic abundance constraint for these choices of parameters, the constraints from direct detection experiments force this mixing angle into a narrow band. The least stringent restriction on \tet{13} is found in the QDF scenario with a DM mass below the top threshold. Since for such a low DM mass, the relic abundance constraint demands a large \lam{3}, we are in the region where the direct detection constraints yield only a mild lower bound on \tet{13}. For larger DM masses, such as in \autoref{Com-deg:sfig2} the third generation coupling has to be smaller and the allowed range for \tet{13} shrinks, small mixing now being excluded. An even more stringent constraint on \tet{13} can be observed in the SFF scenario. Due to the necessary splitting in the couplings, both \lam{1} and \lam{2} are required to be small in the presented cases. Hence the available parameter space is smaller than in the QDF case, so that \tet{13} is more stringently constrained by the LUX data.

The effects on \tet{23} are more subtle but still in some cases visible. In \autoref{Com-deg:sfig1} we see that large values for \tet{23} are slightly preferred in the QDF scenario, yielding the necessary enhancement of the annihilation cros-section. In contrast we can see that in the SFF scenario small values for \tet{23} are preferred, as in this case a reduction of the annihilation cross-section is needed.

We can furthermore observe the already mentioned consequence of demanding a significant mass splitting that leads to SFF in the transition from \autoref{Com-deg:sfig3} to \autoref{Com-deg:sfig4}. Due to the larger DM mass the relic abundance constraint reduces \lam{3} and as a result also \lam{1} and \lam{2}, thereby limiting the possible ranges for the splittings $\Delta_{ij}$. This effect has been discussed in more detail in \autoref{RAsection}.

%\subsection{Impact of Constraints from Future Direct Detection Experiments}

\begin{figure}
  \centering
  \includegraphics[width=.65\linewidth]{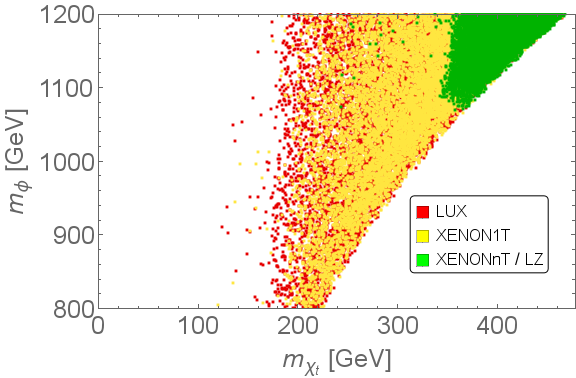}
\caption{Allowed mass ranges for SFF scenario (with $\eta=-0.075$), using expected bounds from future direct detection experiments.}
\label{Mass-BoundsSFF}
\end{figure}

\begin{figure}
  \centering
  \includegraphics[width=.65\linewidth]{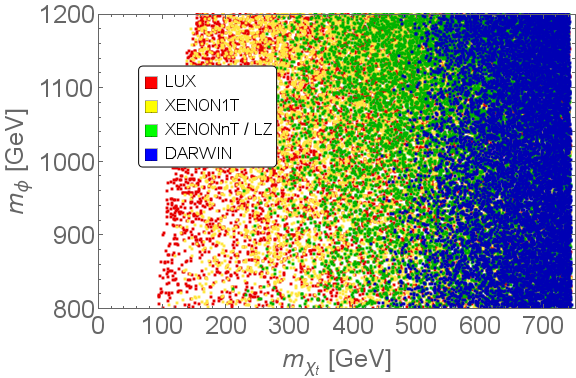}
\caption{Allowed mass ranges for QDF scenario (with $\eta=-0.01$), using expected bounds from future direct detection experiments.}
\label{Mass-BoundsQDF}
\end{figure}

To illustrate the resulting DM mass bounds, \autoref{Mass-BoundsSFF} and \autoref{Mass-BoundsQDF} show the allowed parameter space in the $m_{\chi_t}$-$m_\phi$-plane, again imposing all relevant flavour and DM constraints.
In the SFF scenario, we observe both a lower and an upper bound on the DM mass $m_{\chi_t}$, depending on the mediator mass. We also find that for $m_{\chi_t}<m_{t}$, fine-tuning is required to fulfill all constraints. The observed upper bound depends on the value of $\eta$, a parameter depending on the UV completion of the simplified model considered here. We only display the effects for one specific value, $\eta=-0,075$, but the qualitative effect will remain the same for other values.

In \autoref{XenonIsotopeSection} we found that the expected bounds from future direct detection experiments will exclude large values of \lam3. Together with the constraint from the observed DM relic abundance, which requires (for a fixed mediator mass) larger coupling values (including \lam3) for lower DM masses, this translates into a lower bound on the allowed DM masses. We can see that with improving constraints from direct detection experiments, the lower bound on the DM mass grows. In the SFF scenario, the expected cross-section limit from DARWIN would in fact exclude the whole parameter-space for the SFF scenario (with $\eta=-0,075$). Future direct detection experiments with xenon as well as other materials (with yet different cancellation patterns) are hence crucial to rule out large parts of the parameter space for top-flavoured DM -- or possibly to discover it.

\subsection{Summary of Constraints}\label{subsec:7-summ}

To complete this section, we provide a short summary of the available constraints, as well as an overview of the consequences of their interplay. 

We identified the following crucial constraints on top-flavoured DM beyond MFV:
\begin{itemize}
 \item The  constraints from $D^0-\bar D^0$ mixing requires the mixing angle \tet{12} to be small, unless the couplings \lam{1} and \lam{2} are nearly degenerate.
 \item In order to explain the observed relic abundance by a thermal DM freeze-out,
the DM couplings have to lie in certain ranges (depending on the values of the mediator and DM mass).  
% \item Freeze-out conditions assign valid intervals to \lam{1} and \lam{2}, depending on both \lam{3} and $\eta$.
 \item Due to the large top-quark mass, the mixing angles have an influence on the annihilation cross-section. This effect is sub-leading in the QDF scenario, but significant for SFF -- both \tet{13} and \tet{23} considerably affect the annihilation cross-section. 
 \item Direct detection constraints can only be fulfilled if a near-perfect cancellation of several contributions is realized. For this to happen, either the third generation coupling \lam{3} has to be very small or the mixing \tet{13} has to be in a narrow band. The shape and thickness of this band depends on the other parameters of the model. This destructive interference is required for all isotopes of natural xenon present in the detector. Future direct detection experiments will therefore constrain \lam3 to lower and lower values.
\item
The required destructive interference in the WIMP-nucleus cross-section favours top-flavoured DM over the other cases.
\end{itemize}

Imposing all of the above constraints simultaneously, we discover their non-trivial interplay, leading to the following main conclusions:
\begin{itemize}
 \item For the phenomenologically interesting mass ranges for $\chi$ and $\phi$, the relic abundance constraints demand the coupling \lam3 to lie in a certain range. The direct detection constraints then require \tet{13} to lie in the narrow interval discussed before. The combination of relic abundance and direct detection constraints hence results in stringent bounds for \tet{13}.
 %\item Due to the other constraints involved the ``flexibility'' of the parameters to fulfill a single of these constraints is limited. As a result we observe an even smaller band of valid \tet{13}, being highly preferred by DD constraints. We have seen that the occurrence of this effect depends both on the DM mass and the freeze-out scenario.
 \item In the SFF scenario, the combination of all constraints implies DM masses $m_{\chi_t} < m_{t}$ to be possible only at the price of fine-tuning.
 \item Although we did not discuss the case of two flavours being present at the time of DM freeze-out in this paper, it is straightforward to deduce one consequence of such a scenario. Assuming that we still prefer top-flavoured DM, depending on $\eta$ the dual-flavour freeze-out scenario would demand a splitting between \lam{1} and \lam{2}. Together with the flavour constraints this would result in an upper bound on \tet{12}.
 \item With improving bounds from future direct detection experiments, the combination of direct detection and relic abundance constraints provides increasingly stringent lower bounds on the DM mass (depending on the value of the mediator mass).
\end{itemize}

\section{Summary and Outlook}
\label{summarysection}

In this paper we studied a simplified model of a DM flavour triplet of Dirac fermions coupling to the SM right-handed up-quarks via a new scalar mediator (carrying the gauge charges of the up-quarks). The coupling matrix was left arbitrary, following the principle of Dark Minimal Flavour Violation \cite{DMFVPrimer}. We restricted our attention to the case of top-flavoured DM, which turns out to be phenomenologically preferred.

We started our analysis with estimating the constraints from LHC searches, by comparing the cross-section limits obtained in two representative run 1 searches with the predictions of our model. Assuming a WIMP DM candidate,  $m_\chi \sim  \mathcal{O}(100)$ GeV, we derived a lower bound on the mediator mass in conjunction with an upper bound on the DM-quark couplings. Following this result, we  
restricted the parameter space of the model accordingly for the rest of our analysis.

We then studied in turn the constraints from flavour violating observables, mainly neutral $D$ meson mixing, the constraints from the assumption of DM being a thermal relic, and the bounds on the WIMP-nucleon scattering cross-section implied by direct DM detection experiments. These studies provided interesting and largely complementary constraints on the parameter space of our model. A summary of the obtained constraints can be found in \autoref{subsec:7-summ}.

The most stringent constraints, however, were obtained when taking into account all aforementioned constraints simultaneously, revealing their non-trivial interplay. Particularly the combination of relic abundance and direct detection constraints places strong limits on the model in question. Again, details  can be found in \autoref{subsec:7-summ}. 

We pointed out that the expected improved limits from upcoming direct DM detection experiments will put the scenario of top-flavoured DM under severe pressure. Experiments such as XENON1T, XENONnT, LUX-ZEPLIN (LZ), and DARWIN will hence play an essential role in either ruling out major parts of the parameter space, or in the discovery of top-flavoured DM.

In our analysis we did not include possible constraints from indirect searches for DM, due to the significant uncertainties associated to e.\,g.\ the assumed propagation model. 
Recently however, quite stringent constraints on WIMP DM have been derived \cite{Cuoco:2016eej,Cui:2016ppb} from the latest AMS-02 data \cite{Aguilar:2016kjl}, with the potential to exclude additional parts of the parameter-space.
In our model of top-flavoured DM, we expect the constraints to have a less significant impact than in the generic WIMP case, due to the multiple DM flavours. For instance, in case of multiple flavours being present at the time of freeze-out, the simple relation between the annihilation cross-section relevant for the DM relic abundance and the DM annihilation in our galaxy is lost. In addition, generally the pure top-flavour annihilation cross-section is smaller than the average cross-section, since it is kinematically supressed by the large top quark mass. This in turn suppresses the contribution to indirect detection data. A detailed, quantitative analysis of these constraints is beyond the scope of this paper and left for future research. We note that the hint for WIMP DM with a mass of $\sim80\,\text{GeV}$ \cite{Cuoco:2016eej,Cui:2016ppb} can not be acommodated for in our model, due to the lower bound on the DM mass obtained in \autoref{combinedsection}.

Let us close with a brief comparison of our results to the ones of \cite{DMFVPrimer}, where a simplified model of bottom-flavoured DM was studied. In the latter scenario, flavour constraints from neutral $K$ and
$B$ meson mixings played a crucial role in constraining the flavour mixing angles of the DM-quark coupling matrix. The observed relic abundance, on the other hand, was independent of the amount of flavour mixing. Hence, while in both models strong constraints on the flavour structure of the new coupling matrix were derived, their origins differ from each other. Furthermore, both models are found to be stringently constrained by the absence of signal in direct detection experiments, so that in order to comply with the constraints, a destructive interference between various contributions was required, but the identified cancellation patterns differed from each other. Both models therefore belong to the class of xenophobic DM models\cite{Feng:2013fyw}.

\paragraph{Acknowledgements}

We are grateful to Ulrich Nierste and Jos\'e Zurita for useful discussions.
S.\,K.\ acknowledges the support by the DFG-funded Doctoral School KSETA.

%\phantomsection
\bibliographystyle{JHEP}
\bibliography{sources}
%\addcontentsline{toc}{section}{\numberline{}References}%

\end{document}